# A general approach for high yield fabrication of CMOS compatible all semiconducting carbon nanotube field effect transistors

Muhammad R. Islam[a,b], Kristy J. Kormondy[a,b,], Eliot Silbar[a,b] and Saiful I. Khondaker[a,b,c] *

[a] Nanoscience Technology Center, [b] Department of Physics, [c] School of Electrical Engineering and Computer Science, University of Central Florida, Orlando, Florida 32826, USA.

* To whom correspondence should be addressed. E-mail: saiful@ucf.edu

## ABSTRACT

We report strategies of achieving both high assembly yield of carbon nanotubes at selected position of the circuit via dielectrophoresis (DEP) and field effect transistor (FET) yield using semiconducting enriched single walled carbon nanotube (s-SWNT) aqueous solution. When the DEP parameters were optimized for the assembly of individual s-SWNT, 97% of the devices show FET behavior with a maximum mobility of 210 cm$^2$/Vs, on-off current ratio ~ 10$^6$ and on conductance up to 3 µS, however with an assembly yield of only 33%. As the DEP parameters were optimized so that 1-5 s-SWNTs are connected per electrode pair, the assembly yield was almost 90% with ~ 90% of these assembled devices demonstrating FET behavior. Further optimization gives an assembly yield of 100% with up to 10 SWNT/site, however with a reduced FET yield of 59%. Improved FET performance including higher current on–off ratio and high switching speed were obtained by integrating a local Al$_2$O$_3$ gate to the device. Our 90% FET with 90% assembly yield is the highest reported so far for carbon nanotube devices. Our study provides a pathway which could become a general approach for the high yield fabrication of CMOS compatible carbon nanotube FETs.

## 1. Introduction

Single-walled carbon nanotubes (*SWNTs*) have attracted tremendous attention as a promising building block for future nanoelectronic circuits [1-4]. Field effect transistors (FET) fabricated from individual SWNTs have demonstrated outstanding device performances including high mobility, near ballistic conductance and resistance against electromigration, surpassing the properties of current Si based complimentary metal oxide semiconductor (CMOS) devices [5-10]. However, there are several challenges that need to be addressed before SWNT based electronics can find widespread practical applications. The most challenging among them are the large scale integration of SWNT FETs [4]. This can be addressed (i) by developing strategies to assemble SWNTs at selected positions of the circuits with high yield, and (ii) since only semiconducting SWNT can be used for FET fabrication, it is also necessary to find strategies to separate SWNTs by chirality with a tight control on their diameter. Several techniques have been developed for high yield directed assembly of SWNTs, including direct growth by chemical vapor deposition (CVD) [11-13] and post-growth solution processing by AC dielectrophoresis (DEP) [14-20]. The CVD technique uses patterned catalytic islands in combination with high-temperature treatment to grow nanotubes directly on a substrate [11-13]. The best optimization of assembly and FET yield reported by Javey et al [12] showed that using the direct growth technique it is possible to assemble individual SWNTs with about 40% yield out of which approximately two thirds showed FET behavior, the remaining one third was metallic. Although recent progress has been made in limited chirality control of the SWNTs during direct growth [21-22], the assembly yield were not reported. Furthermore, CVD process requires high growth temperatures in excess of 900 °C making it incompatible with current CMOS technology.

An attractive alternative to direct growth technique for CMOS compatible device integration is the assembly of SWNT from solution phase. Although there are speculations that solution processed SWNTs may be defective and may not be practical for high quality device applications, we have recently shown that progress in solution processing allow us to have SWNTs with low defects which can make high quality devices [17, 23]. Controlled assembly of the solution processed SWNTs have been demonstrated at individual level as well as in an array via DEP [14-20, 24-26]. In DEP, metallic SWNT feel a greater force then semiconducting SWNT during the assembly [27]. Therefore, while using a mixed solution of semiconducting and metallic SWNTs, a FET yield of 50% or lower was demonstrated, leaving a large fraction of devices non-functioning [17-19]. In this respect, it is important to fabricate devices using all semiconducting nanotube solution.

Recently, solution based sorting techniques have been developed to separate nanotubes by chirality [28-35]. In particular, density gradient ultracentrifugation (DGU) approach of surfactant-based separation has been used to sort semiconducting SWNT in aqueous solution [28-31]. Raman studies of such sorted SWNTs show 99% of them are semiconducting [29]. Using these semiconducting-enriched SWNT (s-SWNT), we have recently reported 99% FET yield in a local gated geometry using individual s-SWNT, however with an assembly yield of 20% [36]. Ganzhorn *et al* [37] showed that it is possible to increase the assembly yield to 98% using 1-3 s-SWNT per electrode pair. However, a complete and systematic study regarding FET yield of the assembled devices and variation of FET performance with the number of SWNT is lacking.

In this study, we report systematic investigation and optimization of s-SWNT assembly yield and FET device yield. The s-SWNTs dispersed in water with a diameter distribution of 0.5 to 3.9 nm were assembled between prefabricated taper shaped Pd source and drain electrodes of 1 µm separation via dielectrophoresis (DEP). By applying an ac voltage of 5 Vpp at 1MHz between the electrodes and varying the assembly time from 5-35 sec, we were able to controllably vary the assembly yield from 33% to 100% with the number of s-SWNT varying from 1-10 per electrodes. We then surveyed the electronic properties of the devices statistically. When an assembly time of 5 sec were used, the yield was 33% with the majority of the devices containing individual s-SWNTs. Form the electron transport measurements of these devices we found that 97% show FET behavior (current on off ratio >10). However, the device properties such as mobility, current on-off ratio, on-conductance and subthreshold swing varied over several orders of magnitude form device to device possibly owing to the variation of SWNT diameters. The assembly yield was increased to 90% when a DEP time of 20 sec was used, however each electrode pairs were connected by 1-5 s-SWNTs. Transport measurements revealed that the FET yield slightly decreased to 90%. A 100% assembly yield was obtained by further increasing the DEP time to 35 sec but this increases the number of s-SWNT per device up to 10 and reduces the FET yield to 59%. It was also found that the switching speed of the assembled devices degrade rapidly as the number of SWNT increases. An $Al_2O_3$ local gated structure was used for devices containing 1-5 s-SWNT to improve the switching speed that reduces the subthreshold swing by at least an order of magnitude compared to the back gated device. Our finding of 90% SWNT FET with 90% assembly yield is among the best reported so far and is a major advancement towards the practical realization of SWNT based electronic devices. The approach presented here could become a general approach for the high yield fabrication of SWNT FETs.

## 2. Experimental

The devices were fabricated on heavily doped Si substrates with a 250 nm-capped layer of thermally grown $SiO_2$. Larger size contact pads and position alignment markers were fabricated using optical lithography followed by thermal evaporation of chromium (Cr) (5 nm) and Au (45 nm) and standard lift-off. Arrays of taper-shaped drain and source electrodes with a channel length of 1 µm were defined by electron beam lithography (EBL) using single layer PMMA resists and then developed in (1:3) methyl isobutyl ketone : isopropyl alcohol (MIBK:IPA). After defining the electrode patterns, 3 nm Cr and 25 nm Pd were deposited by electron beam deposition followed by lift off in acetone at 60 °C. The local gate (width ~100 nm) was defined by EBL followed by e-beam evaporation of 3 nm Cr and 20 nm Al with a liquid nitrogen cooled stage. After lift-off, the devices were treated in oxygen plasma to obtain a thin layer of $Al_2O_3$.

The s-SWNT aqueous solution used in this study was obtained from NanoIntegris [31]. The s-SWNTs have a diameter distribution of 0.5 to 3.9 nm with an average of average of 1.6 nm, while the length ranged from 0.5 to 4.0 μm with an average value of 1.8 μm as determined from atomic force microscopy (AFM) and scanning electron microscopy (SEM) study [26]. Figure 1(a) shows the schematic of the experimental set up for the DEP assembly. The concentration of s-SWNT in solution was diluted with deionized (DI) water from the original value of 10 µg/mL down to 10 ng/mL. A small drop (~3µL) of solution was then cast onto the chip and an AC voltage of 5V ($V_{p-p}$) at 1MHz was applied for a fixed time between the source and drain electrodes and then moved to the next electrode pair. Due to the AC voltage, a time averaged DEP force is created which causes the SWNTs to move in a translational motion along the electric field gradient and assemble between the electrodes [27]. DEP can be advantageous over other solution-processed techniques because it allows the materials to directly integrate to prefabricated electrodes at the selected positions of the circuits and does not require post etching or transfer printing. The results of the DEP assembly were examined using a Zeiss Ultra 55 field emission SEM. Before electrical measurement the devices were submerged into DI water to remove surfactant from the surface of the SWNT, and the chip was blown dry using $N_2$ gas.

The detailed electronic transport measurements of all the assembled devices were carried out before and after thermal anneal. Electrical measurements were performed in a probe station at ambient condition using a Keithley 2400 source meter and a DL instruments 1211 current preamplifier interfaced with LabView program. After measuring the as-assembled devices, the samples were annealed at 200°C for 1 h in an argon (Ar) /hydrogen ($H_2$) atmosphere and the electrical transport measurements were repeated.

### 3. Results and Discussions:

At first, we optimized the DEP parameters to integrate individual s-SWNTs in the circuit. DEP assembly mainly depends on four parameters: applied AC voltage, sinusoidal frequency, s-SWNT solution concentration and DEP assembly time. For this purpose, an AC voltage of 5V ($V_{p-p}$) at 1MHz was applied for 5 seconds. Out of the 384 electrode pairs used in this study, we found that on an average 25% were bridged by

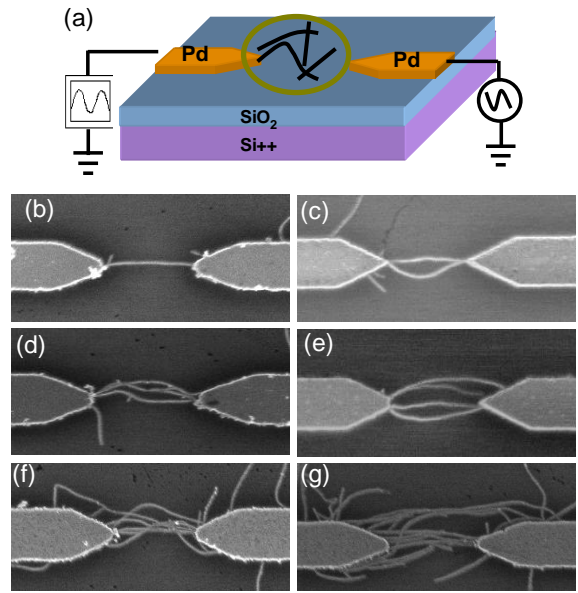

**Fig.1** (a) Schematic diagram of the DEP assembly set up. SEM images for the DEP assembled devices containing (b) one (c) two (d) three (e) four (f) six and (g) eight s-SWNTs between the electrodes..Separation between the electrodes in each case is 1μm individual s-SWNT, another 5% were connected by 2 s-SWNTs and 3% were connected by 3 s-SWNTs giving a total of 33% assembly yield. To improve the assembly yield, the DEP time was increased to 20 sec while keeping other parameters fixed. A total of 76 electrode pairs that we used for this assembly, we found that 69 electrode pairs were bridged with 1-5 s-SWNTs giving an average assembly yield of 90%. In this assembly 15% of the electrodes were connected by individual s-SWNT, 60% by 2-3 s-SWNTs and 25% by 4-5 s-SWNTs. We have attempted to increase the assembly yield further by increasing the DEP assembly time to 35 sec while keeping all other DEP parameters fixed. Out of the 54 electrode pairs used in this study, all of them were connected by 1-10 s-SWNTs per site giving an assembly yield of 100%. Here, 55% of the devices were bridged with 1-5 s-SWNT and the rest of the devices had 6-10 s-SWNT per site. Figure 1 (b-g) show representative SEM images of the DEP assembled s-SWNT devices containing one (fig 1b), two (fig 1c), three (fig 1d), four (fig 1e), six (fig 1f) and eight (fig 1g) s-SWNT per site. The detailed electronic transport measurements of all the assembled devices were then carried out before and after thermal anneal.

Figure 2(a) shows the transfer characteristics (drain current $I_{DS}$ versus backgate voltage $V_G$) measured at a fixed source-drain bias voltage $V_{DS}$ = -0.5V for one of our best performing individual s-SWNT devices before and after thermal annealing. The Si substrate was used as a backgate. The $I_{DS}$ decreased by several orders of magnitude with increasing $V_G$, demonstrating a p-type FET behavior. In addition, the current on-off ($I_{on}/I_{off}$) ratio and on-state conductance ($G_{on} = I_{on}/V_{DS}$) were increased significantly after annealing. The $G_{on}$ was 0.2 µS for the as-assembled device which after thermal annealing increased by an order of magnitude to ~ 2 µS. The $I_{on}/I_{off}$ of the device was increased from $2.3 \times 10^3$ to $2 \times 10^4$ after annealing. The high values of $G_{on}$ and $I_{on}/I_{off}$ are indicative of high quality SWNT FET desvice. We also calculated the subthreshold

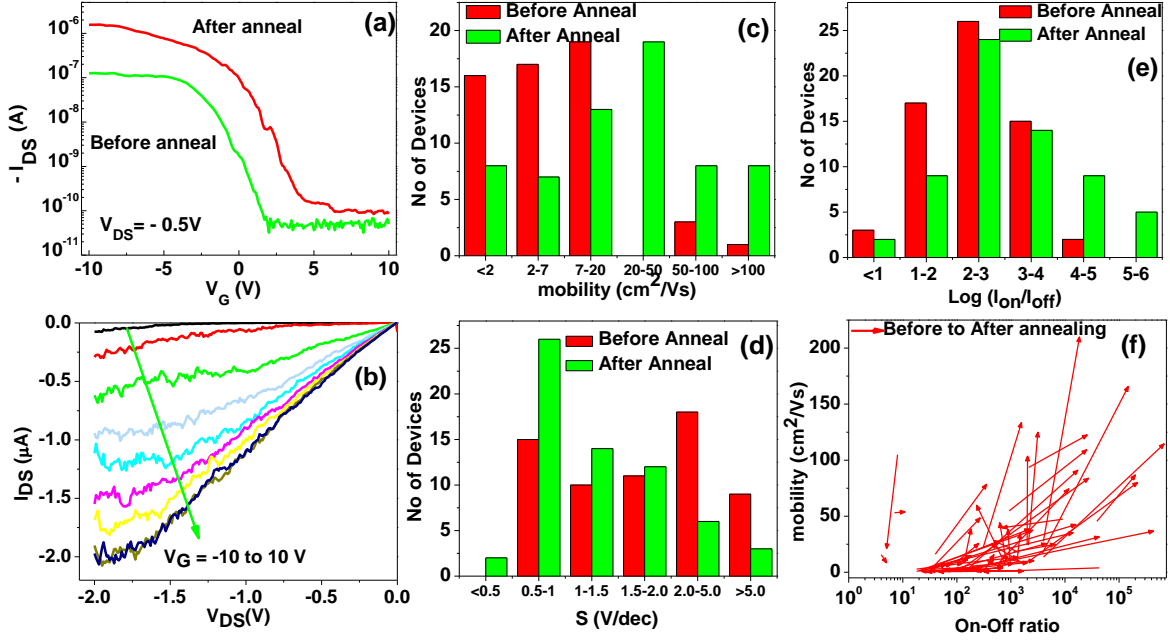

**Fig. 2** (a) Transfer characteristics of a representative individual SWNT FET device before and after annealing. $V_{DS}$ = -0.5 V. (b) Output characteristics of the same device. (c-f) Summary of FET properties for all individual s-SWNT devices before and after anneal. Histogram of mobility(c), subthreshol swing S (d), $I_{on}/I_{off}$ (e), and Vector diagram of mobility and $I_{on}/I_{off}$ representing the changes of each SWNT FET upon annealing (f).

swing $S = [dlogI_D/dV_G]^{-1}$ of the device which were 1.5 V/dec and 1.4 V/dec before and after annealing respectively, typical of a backgated SWNT FET [14-15]. Figure 2(b) shows the output characteristics ($I_D$ vs $V_{DS}$) for the same device up to the saturation regime at different gate voltages (from -10 V to 10 V in a gate steps of 2.5 V) after anneal. The output current show linear behavior at low bias and saturation at high bias voltage indicating that the current is not limited by contact resistance or short channel effect. The linear mobility of the device was calculated using the relation $\mu = (L^2/C_G \times V_{DS}) (dI_{DS}/dV_G)$, where $L$ is the channel length and $C_G = (2\pi\varepsilon L)/ln(2t_{ox}/r)$ is the gate capacitance, with $\varepsilon \sim 3.9\varepsilon_0$ is the effective dielectric constant of $SiO_2$, $h$ is the thickness of the oxide, and $r$ is the radius of the s-SWNT [38]. By considering r = 0.75 nm (measured from AFM) we calculated $\mu$ to be 28 cm$^2$/Vs and 210 cm$^2$/Vs before and after annealing respectively. The value of $\mu$ after anneal is comparable to other high quality solution processed individual s-SWNT FET devices fabricated from mixed SWNT solution [18, 39, 40].

This indicates that the chirally separated s-SWNT used in this study were also of high quality. The increased on current, $I_{on}/I_{off}$ and mobility after annealing can be attributed to the removal of surfactant molecule from the sidewall of the s-SWNT [41] and the subsequent reduction of Schottky barrier height between the Pd electrode and s-SWNT [42]. This reduced barrier favors injection of hole, causing high on current, which in turn increases $I_{on}/I_{off}$ and mobility.

Detailed FET characteristics were measured before and after anneal for a total of 63 individual s-SWNT devices. This is summarized in Figure 2(c-f) where we show histograms of $\mu$, S, $I_{on}/I_{off}$, as well as a vector plot of $\mu$ vs $I_{on}/I_{off}$. Figure 2(c) shows that $\mu$ of the as assembled devices varied from 1 to 113 cm$^2$/Vs with an average of 13 cm$^2$/Vs, which after annealing increased to 2 - 210 cm$^2$/Vs with an average of 40 cm$^2$/Vs. The variation of $S$ before and after annealing is shown in figure 2(d). The median value of S decreased from 1.2 to 1 V/dec after annealing. This improvement in S can be attributed to the reduction of trap charges between the SiO$_2$ and s-SWNT interface [43]. Figure 2(e) shows the median value of $I_{on}/I_{off}$ increased from 184 to 720 (mean value from $2\times10^4$ to $3\times10^5$) upon annealing. The median $G_{on}$ also increased from 0.03 to 0.2 µS due to annealing. Figure 2(f) shows a vector plot of how $\mu$ and $I_{on}/I_{off}$ changes for each of the measured devices upon annealing. A maximum increase of $\mu$ and $I_{on}/I_{off}$ from 28 to 210 and $4\times10^3$ to $6.8\times10^5$ has been observed respectively. As discussed in previous section the reason for such improvements is the reduction of contact barrier upon annealing.

Two important conclusions can be drawn from here. From figure 2(f), we see that out of the 63 measured devices, 61 shows $I_{on}/I_{off}$ more than one order of magnitude, signifying a FET yield of 97%. By combining properties of a few other devices containing 2-3 SWNTs (discussed in the following section) from the 5 sec assembly, the FET yield remained 97%. Here the devices with $I_{on}/I_{off}$ greater than 10 were considered as FET according to ref [44]. However, if we choose the $I_{on}/I_{off}$ >3 as FET as mentioned in ref [19] the FET yield would become 99%. Figure 3 also show that the FET properties vary from device to device. Such variations may be explained by the differences of chirality and diameter from tube to tube; these physical features affect the bandgap and contact resistance of the s-SWNT, and in turn the mobility, on/off current ratio and subthreshold swing of each device [45-48]. This means that even for all s-SWNT characterized as semiconducting, narrowing the distribution of diameters and chirality may facilitate more homogeneous device behavior and future effort should be directed in improved sorting of s-SWNTs by diameter. Such device to device variation may also explain many interesting behavior, such as the tradeoff between $\mu$ and $I_{on}/I_{off}$, observed in semiconducting rich nanotube networks and arrays [26, 49-52].

We now examine the transport characteristics of the devices assembled with a DEP time of 20 sec containing 1-5 s-SWNTs. Since the annealed devices show better performance compared to the as-assembled devices as discussed in previous section, from now on we only discuss the FET properties of annealed devices. Figure 3(a) show the transfer characteristics of a representative device with 2 s-SWNT/site showing a p-type FET behavior with an on-conductance of 3.4 µS and $I_{on}/I_{off}$ ~$2\times10^4$. Figure 3(b) show the output characteristic of this device up to the saturation regime at different gate voltages with an output current of up to 8µA. The value of $I_{on}/I_{off}$ for this device is similar to that of our best individual s-SWNT, however the on-conductance and output current is higher than the individual SWNT. For devices containing more than 1 SWNT, we approximated the mobility using the SWNT array formula $\mu=(L/WC_iV_{DS})(dI_{DS}/dV_G)$, where, $D$ is the linear density of the SWNT, $W$ is the channel width, and $C_i = D/[C_Q^{-1} + (1/2\pi\varepsilon)\ln[\sinh(2\pi t_{ox}D)/\pi Dr]]$, is the specific capacitance per unit area of aligned array with $C_Q$ (= $4\times10^{-10}$ F/m) is the quantum capacitance of nanotube [24, 26]. For 2 s-

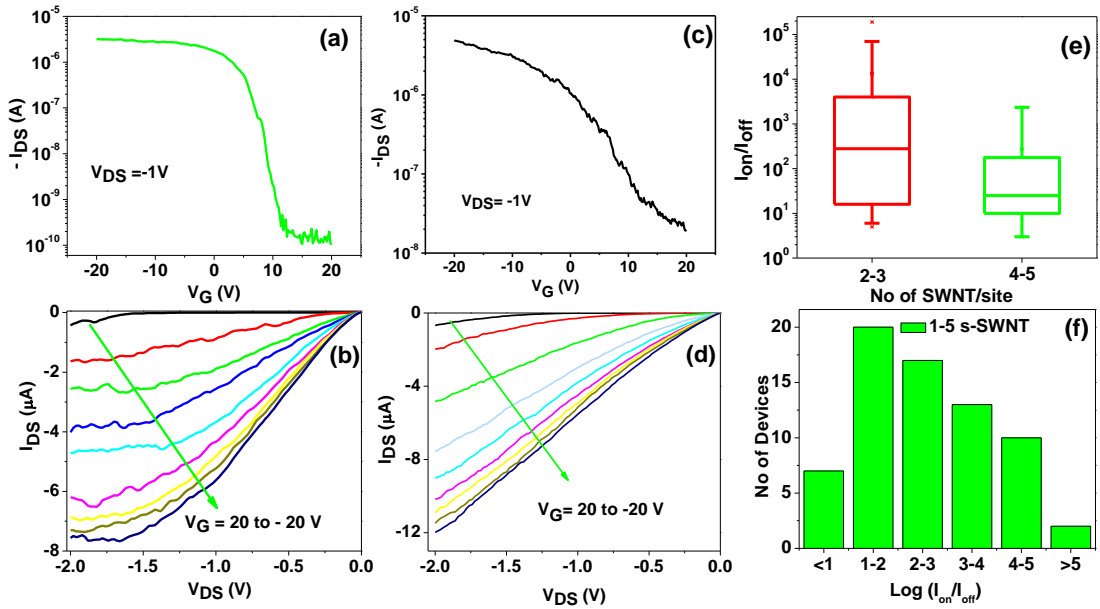

**Fig. 3** (a) Transfer characteristics of a FET device with 2 s-SWNT. (b) output characteristics of the same device for different gate voltages ranging from +20 V to -20 V. (c) Transfer and (d) and output characteristics of a device with 4 s-SWNT. (e) Box plot of $I_{on}/I_{off}$ for 2-3 and 4-5 s-SWNT showing the range and median value. (f) Histogram of the $I_{on}/I_{off}$ for all devices with 1-5 s-SWNT obtained from 20 sec assembly.

SWNT/site, we estimated an upper limit of $W = 0.25$ µm and D=8 from the SEM image of Fig 1c, and calculated µ=126 cm$^2$/Vs, close to the µ value of our best individual SWNT device. The value of S = 1.9 V/dec is also similar to what has been obtained for individual s-SWNT devices. Similar FET characteristics were measured for many devices containing 2-3 s-SWNT. The $I_{on}/I_{off}$ for 41 such devices are shown in the box diagram (red box) of figure 3(e). From here, we see that the $I_{on}/I_{off}$ of most of the devices bridged by 2-3 s-SWNT are higher than two orders of magnitude with a median value ~3×10$^2$, similar to what has been observed for the individual s-SWNT device. Out of the 41 devices, 38 showed $I_{on}/I_{off}$ more than one order of magnitude. These results suggest that FETs fabricated with 2-3 s-SWNTs are similar to that of the individual s-SWNT FET, however with an increased on conductance and output current. Transfer characteristics of another device with 4 s-SWNT is shown in figure 3(c) with an $I_{on}/I_{off}$ ~10$^2$. Output characteristic of the same device is presented in figure 3(d), showing a higher output current of 12 µA, however no saturation was observed. The linear mobility of the device was calculated to be 134 cm$^2$/Vs using $W = 0.5$µm (see Fig 1e) and D=8. The green box of figure 3(e) represents the variation of $I_{on}/I_{off}$ of all the 18 devices containing 4-5 s-SWNT. We observed that the median $I_{on}/I_{off}$ for 4-5 s-SWNT devices were 25, one order of magnitude lower compared to 1-3 s-SWNT devices. This may be explained using figure 2(f) where we demonstrated a variation of $I_{on}/I_{off}$ from 10 to $10^5$ for the individual s-SWNT FETs. When there are more than 1 s-SWNTs per site, the chances of getting one larger diameter s-SWNT per site increases. Since the total $I_{on}/I_{off}$ is limited by the s-SWNT with lower $I_{on}/I_{off}$, the overall $I_{on}/I_{off}$ of the device decreases. In addition, when the number of SWNT per site increases, the inter-tube spacing also decreases. If this spacing is lower than the gate oxide (SiO$_2$) thickness of 250 nm, the gate voltage may be screened causing an incomplete depletion of charge carriers with gate voltage, resulting in a lower $I_{on}/I_{off}$ [49, 53]. Figure 3(f) represents an overall summary of the $I_{on}/I_{off}$ of all the s-SWNT device with 1-5 s-SWNT per site resulting from 20 sec DEP assembly. From here, we see that out of the 69 devices measured, 62 showed $I_{on}/I_{off} > 10$ giving a 90% FET yield. It is important to note that such a high FET yield cannot be obtained with mixed nanotube solution since increasing the number of SWNTs per site will inevitably result in an increased possibility of at least one metallic SWNT per site.

Transport properties were also studied for the 54 devices assembled with a DEP time of 35 sec containing 1-10 s-SWNTs with an assembly yield of 100%. Figure 4(a) shows the transfer characteristics of a representative device with 6 s-SWNTs/site. This device has an $I_{on}/I_{off}$ of only 20 with µ=87 cm$^2$/Vs. A very high on conductance of ~7µS was observed which is expected as the on current is the sum of the on current of individual SWNT. The devices consisting of 8-10 s-SWNT shows almost metallic behavior with $I_{on}/I_{off}$ <10. Figure 4(b) shows the histogram of $I_{on}/I_{off}$ determined from the transport measurements of the 54 devices. Only 59% of the devices show FET behavior, a significant decrease compared to 1-5 s-SWNT devices.

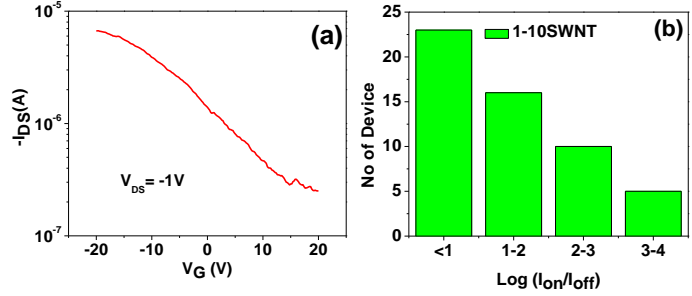

**Fig. 4** (a) Transfer characteristics of a represntative FET device with 6 s-SWNT. (b) Histogram of the $I_{on}/I_{off}$ for device with 1-10 s-SWNT devices.

Table 1 summarizes the assembly yield from a total of 504 electrode pairs used in this study for different DEP assembly time. It also summarizes corresponding FET yield from the electrical measurements of 186 devices. A higher FET yield (97%) was obtained with 1-3 s-SWNT per site using low DEP time ( 5 sec), although this assembly had almost two third of the electrode pairs empty. Whereas using a higher DEP time (35 sec) it was possible to obtain 100% assembly yield, however, the FET yield was 59%. Higher assembly yield (90%) with high FET yield (90%) was obtained using an intermediate DEP time of 20 sec, where the device consists of 1-5 s-SWNT per site. Our 90% FET yield with 90% assembly yield is the highest reported so far for carbon nanotube devices. We note that if the FET is defined with $I_{on}/I_{off} > 3$ the corresponding FET yield will be 99%, 94% and 73% for 5, 20 and 35 sec assembly respectively.

TABLE 1. Summary of the assembly and FET yield of the s-SWNT device fabricated using different DEP assembly times.

| Assembly Time (sec) | Assembly Yield | No of SWNT/ site | FET yield | |
|---|---|---|---|---|
| | | | $I_{on}/I_{off} > 10$ | $I_{on}/I_{off} > 3$ |
| 5 | 33% | 1-3 | 97% | 99% |
| 20 | 90% | 1-5 | 90% | 94% |
| 35 | 100 | 1-10 | 59% | 73% |

We now turn to the switching speed of the fabricated devices. The sub-threshold swing S is a key parameter for any FET and low value of S is necessary for faster switching and low power operation. The theoretical limit of S for SWNT FET is 60 mV/dec [7]. Figure 5 show how the median value of S for our backgated devices ($S_{BG}$) vary with the number of s-SWNT per site. For individual s-SWNT, $S_{BG}$ = 2 V/dec, which slowly increased to 4.2 V/dec for 3 s-SWNT/site. It then rapidly increased to 40 V/dec for 6-10 s-SWNT/site, much higher compared to the theoretical value. The value of S depends on the ratio of the gate capacitance ($C_G$) to the capacitance resulting from charges trapped at the interface between gate oxide and SWNTs ($C_T$) and follow the relation, $S = (2.303 K_B T/e)(C_T/C_G)$ where, $T$ is absolute temperature, $K_B$ is Boltzman constant and $e$ is electron charge [54]. As the number of s-SWNT per device increased, the trapped charges between the s-SWNT and gate oxide increases causing higher $C_T$, this may be one of the reasons for the observed large value of S. Additionally, with a thick (250 nm) layer of oxide ($SiO_2$) the gate coupling becomes less efficient as the number of s-SWNT/site increases, resulting in a higher $S_{BG}$ (slower switching speed).

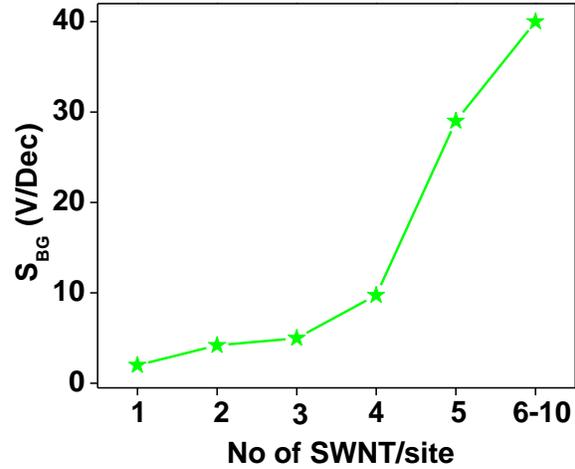

**Fig. 5** Variation of the median value of subthreshold swing $S_{BG}$ for the back gated devices with the number of SWNT/site

To obtain a better switching performance, we have fabricated FET devices with a local Al/Al$_2$O$_3$ bottom gate with a gate oxide thickness of 2-3 nm containing 1-5 s-SWNT between the source and drain electrodes (20 sec assembly). The assembly of s-SWNT was not influenced by the presence of the local gate and the assembly yield was consistent with the aforementioned

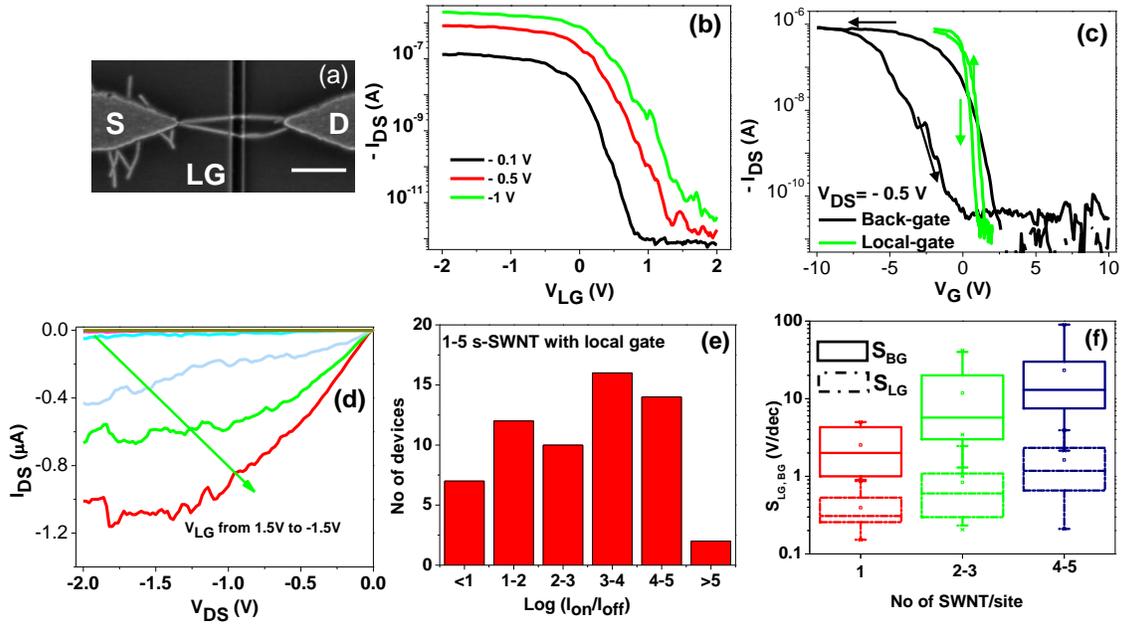

**Fig. 6** (a) Representative SEM image of a device with 2 s-SWNT between the source and drain in a local gated (LG) geometry. Scale bar: 500 nm. (b) Transfer characteristics of a local gated FET device with 2 s-SWNT (c) comparison of back gate ( black curve) and local gate (green curve) The back gated device shows large hysteresis whereas the local gated device shows small hysteresis. (d) Output characteristics of the same device for different local gate voltage. (e) Histogram of the $I_{on}/I_{off}$ for all local gated devices with 1-5 s-SWNT, and (f) Box plot showing variation of subthreshold swing with the number of s-SWNT for back gate ($S_{BG}$) and local gate ($S_{LG}$) devices.

back-gated devices. A representative local gated device with 2 s-SWNT is shown in figure 6(a). The transfer characteristics of the device at different bias voltages are shown in figure 6(b). It shows high $I_{on}/I_{off} \sim 6\times10^5$ at $V_{DS}$ = - 0.5V. The gate leakage current measured for the local gated device is negligible (< 1 pA) for the gate voltage range used. The subthreshold swing for this local gated device ($S_{LG}$) was 245 mV/dec. Figure 6(c) shows the transfer characteristics of the same device for the back and local gated operation for comparison. The $S_{BG}$ measured for this device is 1.6 V/dec nearly one order of magnitude higher than $S_{LG}$. In addition, the local gate operation results in a much reduced hysteresis (~ 0.4 V) compared to the back gated operation (~3.5 V). The large value of hysteresis with back gate can be attributed to the charge trapped in bulk $SiO_2$ or in the Si/ $SiO_2$ interface, which is mainly due to the presence of water molecule in the surface [43]. The much thinner oxide layer and reduced gate length of the local gated geometry causes less charge trapping resulting in reduced hysteresis. Figure 6(d) show the output characteristics at different local gate voltage for this device. The device shows a good gate modulation exhibiting linear region at low bias and saturation region at high bias voltage.

Electron transport properties of 58 local gated devices with 1-5 s-SWNT/site were studied and the histogram for the $I_{on}/I_{off}$ of the devices is shown in figure 6(e). Out of them, 52 devices show $I_{on}/I_{off}$ >10, signifying a 90% FET yield similar to what has been observed for the back gated structure. The median value of $I_{on}/I_{off}$ ~1.5×10$^3$, higher compared to that of the back gated device due to a better gate coupling. Figure 6(f) shows the value of $S_{LG}$ plotted versus the number of s-SWNT/site along with $S_{BG}$ for comparison. The median value of $S_{LG}$ was 300, 640 and 1170 mV/dec for individual, 2-3 SWNT/site and 4-5 SWNT/site respectively. These values are at least one order of magnitude lower than $S_{BG}$. The best value of $S_{LG}$ = 140 mV/dec was obtained from an individual s-SWNT device. The reduction in S with local gate can be attributed to the decrease of $C_T/C_G$ ratio with Al$_2$O$_3$ dielectric. Considering $t_{ox}$=3 nm and $\varepsilon$~10 for Al$_2$O$_3$, we can approximate the ratio of gate capacitance for Al$_2$O$_3$ : SiO$_2$ = 10:1. This matches closely with the result presented in figure 6(f). Therefore, a better FET performance with high $I_{on}/I_{off}$, low S and reduced hysteresis, along with high FET yield was achieved by integrating s-SWNT in a local bottom gated structure.

## 4. Conclusion

In summary, we presented a tradeoff between high SWNT FET yield and high assembly yield using solution processed s-SWNT. The controlled assembly at the selected position of the circuit was done via DEP and electron transport properties of the fabricated devices were investigated. We showed that individual s-SWNT devices gives high quality FET performance with a 97% FET yield but with a low assembly yield (33%). By optimization of the DEP parameters, we obtained a 90% assembly yield with 1-5 s-SWNT per electrode out of which 90% of the devices showed FET behavior, which is the highest reported so far. An improved FET performance including reduced hysteresis and faster switching speed was demonstrated by integrating a local $Al_2O_3$ gates. The high-yield FET fabrication technique using all semiconducting carbon nanotubes demonstrated here is a significant step forward for the practical realization of SWNT based nanoelectronic devices.

**Acknowledgment.** This work is partially supported by the U.S. National Science Foundation under Grant ECCS-0748091 (CAREER).